\documentstyle[twoside,epsf]{article}

\catcode`\@=11
\long\def\@makefntext#1{
\protect\noindent \hbox to 3.2pt {\hskip-.9pt  
$^{{\eightrm\@thefnmark}}$\hfil}#1\hfill}       

\def\@makefnmark{\hbox to 0pt{$^{\@thefnmark}$\hss}}    
    
\def\ps@myheadings{\let\@mkboth\@gobbletwo
\def\@oddhead{\hbox{}
\rightmark\hfil\eightrm\thepage}   
\def\@oddfoot{}\def\@evenhead{\eightrm\thepage\hfil
\leftmark\hbox{}}\def\@evenfoot{}
\def\sectionmark##1{}\def\subsectionmark##1{}}

\oddsidemargin=\evensidemargin
\newcounter{sectionc}\newcounter{subsectionc}\newcounter{subsubsectionc}
\renewcommand{\section}[1] {\vspace{12pt}\addtocounter{sectionc}{1} 
\setcounter{subsectionc}{0}\setcounter{subsubsectionc}{0}\noindent 
    {\tenbf\thesectionc. #1}\par\vspace{5pt}}
\renewcommand{\subsection}[1] {\vspace{12pt}\addtocounter{subsectionc}{1} 
\setcounter{subsubsectionc}{0}\noindent 
{\bf\thesectionc.\thesubsectionc. {\kern1pt \bfit #1}}\par\vspace{5pt}}
\renewcommand{\subsubsection}[1] {\vspace{12pt}\addtocounter{subsubsectionc}{1}
    \noindent{\tenrm\thesectionc.\thesubsectionc.\thesubsubsectionc.
    {\kern1pt \tenit #1}}\par\vspace{5pt}}
\newcommand{\nonumsection}[1] {\vspace{12pt}\noindent{\tenbf #1}
    \par\vspace{5pt}}

\newcounter{appendixc}
\newcounter{subappendixc}[appendixc]
\newcounter{subsubappendixc}[subappendixc]
\renewcommand{\thesubappendixc}{\Alph{appendixc}.\arabic{subappendixc}}
\renewcommand{\thesubsubappendixc}
    {\Alph{appendixc}.\arabic{subappendixc}.\arabic{subsubappendixc}}

\renewcommand{\appendix}[1] {\vspace{12pt}
        \refstepcounter{appendixc}
        \setcounter{figure}{0}
        \setcounter{table}{0}
        \setcounter{lemma}{0}
        \setcounter{theorem}{0}
        \setcounter{corollary}{0}
        \setcounter{definition}{0}
        \setcounter{equation}{0}
        \renewcommand{\thefigure}{\Alph{appendixc}.\arabic{figure}}
        \renewcommand{\thetable}{\Alph{appendixc}.\arabic{table}}
        \renewcommand{\theappendixc}{\Alph{appendixc}}
        \renewcommand{\thelemma}{\Alph{appendixc}.\arabic{lemma}}
        \renewcommand{\thetheorem}{\Alph{appendixc}.\arabic{theorem}}
        \renewcommand{\thedefinition}{\Alph{appendixc}.\arabic{definition}}
        \renewcommand{\thecorollary}{\Alph{appendixc}.\arabic{corollary}}
        \renewcommand{\theequation}{\Alph{appendixc}.\arabic{equation}}
        \noindent{\tenbf Appendix \theappendixc #1}\par\vspace{5pt}}
\newcommand{\subappendix}[1] {\vspace{12pt}
        \refstepcounter{subappendixc}
        \noindent{\bf Appendix \thesubappendixc. {\kern1pt \bfit #1}}
    \par\vspace{5pt}}
\newcommand{\subsubappendix}[1] {\vspace{12pt}
        \refstepcounter{subsubappendixc}
        \noindent{\rm Appendix \thesubsubappendixc. {\kern1pt \tenit #1}}
    \par\vspace{5pt}}

\newcommand{\textlineskip}{\baselineskip=13pt}
\newcommand{\smalllineskip}{\baselineskip=10pt}

\newcommand{\copyrightheading}[1]
    {\vspace*{-2.5cm}\smalllineskip{\flushleft
    {\footnotesize Quantum Information and Computation, Vol.~1, No.~0 (2001) 000--000 #1}\\
    {\footnotesize \copyright\kern2pt Rinton Press}\\
     }}

\renewenvironment{thebibliography}[1]
        {\frenchspacing
     \ninerm\baselineskip=11pt
         \begin{list}{\arabic{enumi}.}
        {\usecounter{enumi}\setlength{\parsep}{0pt}     
     \setlength{\leftmargin 12.7pt}{\rightmargin 0pt}
         \setlength{\itemsep}{0pt} \settowidth
    {\labelwidth}{#1.}\sloppy}}{\end{list}}

\newcounter{itemlistc}
\newcounter{romanlistc}
\newcounter{alphlistc}
\newcounter{arabiclistc}

\newcommand{\fcaption}[1]{
        \refstepcounter{figure}
        \setbox\@tempboxa = \hbox{\footnotesize Fig.~\thefigure. #1}
        \ifdim \wd\@tempboxa > 5in
           {\begin{center}
        \parbox{5in}{\footnotesize\smalllineskip Fig.~\thefigure. #1}
            \end{center}}
        \else
             {\begin{center}
             {\footnotesize Fig.~\thefigure. #1}
              \end{center}}
        \fi}

\newcommand{\tcaption}[1]{
        \refstepcounter{table}
        \setbox\@tempboxa = \hbox{\footnotesize Table~\thetable. #1}
        \ifdim \wd\@tempboxa > 5in
           {\begin{center}
        \parbox{5in}{\footnotesize\smalllineskip Table~\thetable. #1}
            \end{center}}
        \else
             {\begin{center}
             {\footnotesize Table~\thetable. #1}
              \end{center}}
        \fi}

\def\pmb#1{\setbox0=\hbox{#1}
    \kern-.025em\copy0\kern-\wd0
    \kern.05em\copy0\kern-\wd0
    \kern-.025em\raise.0433em\box0}

\def\fnt#1#2{\footnotetext{\kern-.3em
    {$^{\mbox{\scriptsize #1}}$}{#2}}}

\def\fpage#1{\begingroup
\voffset=.3in
\thispagestyle{empty}\begin{table}[b]\centerline{\footnotesize #1}
    \end{table}\endgroup}

\def\runninghead#1#2{\pagestyle{myheadings}
\markboth{{\protect\footnotesize\it{\quad #1}}\hfill}
{\hfill{\protect\footnotesize\it{#2\quad}}}}
\font\tenrm=cmr10
\font\tenit=cmti10 
\font\tenbf=cmbx10
\font\bfit=cmbxti10 at 10pt
\font\ninerm=cmr9

\font\eightrm=cmr8

\newtheorem{proposition}{Proposition}

\def\FigName{figure}%
\newbox\captionbox
\long\def\@makecaption#1#2{%
  \ifx\FigName\@captype
    \vskip\abovecaptionskip
    \setbox\tempbox\hbox{{\figurecaptionfont #1\hskip1em #2}}
    \ifdim\wd\tempbox< 28pc
    \centerline{\box\tempbox}
    \else
    {\figurecaptionfont #1\hskip1em #2\par}
\fi\else
    \setbox\tempbox\hbox{{\tablecaptionfont #1\hskip1em #2}}
    \ifdim\wd\tempbox< 28pc 
    \centerline{\box\tempbox}
    \else
    {\tablecaptionfont #1\hskip1em #2\par}%
    \fi   
 \vskip\belowcaptionskip
 \fi}
\InputIfFileExists{psfig.sty}
{\typeout{^^Jpsfig.sty inputed...ok}}{\typeout{^^JWarning: psfig.sty could be be found.^^J}}
\InputIfFileExists{epsfsafe.tex}
{\typeout{^^Jepsfsafe.tex inputed...ok}}
            {\typeout{^^JWarning: epsfsafe.tex could not be found.^^J}}
\InputIfFileExists{epsfig.sty}
{\typeout{^^Jepsfig.sty inputed...ok}}{\typeout{^^JWarning: epsfig.sty could not be found.^^J}}
\InputIfFileExists{epsf.sty}
{\typeout{^^Jepsf.sty inputed...ok}}{\typeout{^^JWarning: epsf.sty could not be found.^^J}}%
\def\fps@figure{tbp}
\def\ftype@figure{1}
\def\ext@figure{lof}
\def\fnum@figure{Fig.\ \thefigure}
\def\qed{\hbox{${\vcenter{\vbox{              
   \hrule height 0.4pt\hbox{\vrule width 0.4pt height 6pt
   \kern5pt\vrule width 0.4pt}\hrule height 0.4pt}}}$}}

\begin{document}
\setlength{\textheight}{7.7truein}    

\runninghead{Book review 
}{}          
\normalsize\textlineskip
\thispagestyle{empty}
\setcounter{page}{95}

\copyrightheading{} 

\vspace*{0.88truein}

\fpage{95}
\centerline{\bf
BOOK REVIEW}
\vspace*{0.15truein}
\centerline{ \bf on }
\vspace*{0.37truein}
\centerline{
\bf A New Kind of Science }
\vspace*{0.015truein}
\centerline{
by Stephen Wolfram}
\vspace*{0.015truein}
\centerline{
\it Wolfram Media, Inc., May 2002}
\vspace*{0.015truein}
\centerline{\it Hardcover \$44.95 (1192 pages)  ISBN: 1579550088
 }
\vspace*{0.035truein}

\vspace*{0.21truein}

\vspace*{10pt}

\begin{quote}
\textquotedblleft Somebody says, `You know, you people always say that space
is continuous. How do you know when you get to a small enough dimension that
there really are enough points in between, that it isn't just a lot of dots
separated by little distances?' Or they say, `You know those quantum
mechanical amplitudes you told me about, they're so complicated and absurd,
what makes you think those are right? Maybe they aren't right.' Such remarks
are obvious and are perfectly clear to anybody who is working on this problem.
It does not do any good to point this out.\textquotedblright

---Richard Feynman \cite[p.161]{feynman}
\end{quote}

\vspace*{1pt}\textlineskip
\section{Introduction}
\vspace*{-0.5pt}
\noindent

\textit{A New Kind of Science} \cite{wolfram}, the 1280-page treatise by
\textit{Mathematica} creator Stephen Wolfram, has only a few things to say
about quantum computing. Yet the book's goal---to understand nature in
computational terms---is one widely shared by the quantum computing community.
\ Thus, many in the field will likely be curious: is this 2.5-kilogram tome
worth reading? \ Notwithstanding newspaper comparisons \cite{farmela} to
Darwin's \textit{Origin of Species}, what is the book's actual content? \ This
review will not attempt a chapter-by-chapter evaluation, but will focus on two
areas: computational complexity and fundamental physics.

As a popularization, \textit{A New Kind of Science} is an impressive
accomplishment. \ The book's main theme is that simple programs can exhibit
complex behavior. \ For example, let $p_{i,j}=1$ if cell $(i,j)$ is colored
black, and $p_{i,j}=0$ if white. \ Then the `Rule $110$' cellular automaton is
defined by the recurrence
\[
p_{i+1,j}=p_{i,j}+p_{i,j+1}-\left(  1+p_{i,j-1}\right)  p_{i,j}p_{i,j+1}%
\]
for $i\geq0$, given some initial condition at $i=0$. \ Wolfram emphasizes that
such an automaton, even when run with a simple initial condition such as a
single black cell, can generate a complicated-looking image with no apparent
repetitive or nested structure. \ This phenomenon, although well known to
programming enthusiasts as well as professionals, will no doubt surprise many
general readers.

Using cellular automata as a framework, Wolfram moves on to discuss a range of
topics---including the second law of thermodynamics, natural selection, plant
and animal morphology, artificial intelligence, fluid dynamics, special and
general relativity, quantum mechanics, efficient algorithms and \textsf{NP}%
-completeness, heuristic search methods, cryptography and pseudorandomness,
data compression, statistical hypothesis testing, G\"{o}del's Theorem,
axiomatic set theory, and the Church-Turing thesis. \ What is noteworthy is
that he explains all of these without using formal notation. \ To do so, he
relies on about $1000$ high-resolution graphics, which often (though not
always) convey the ideas with as much precision as a formula would. \ With
suitable disclaimers, \textit{A New Kind of Science} could form an excellent
basis for an undergraduate general-science course.

The trouble is that, as the title implies, Wolfram emphatically does not
believe that he is using cellular automata to popularize known ideas. \ In the
introduction, he describes his finding that one-dimensional cellular automata
can produce complex behavior as one of the ``more important single discoveries
in the whole history of theoretical science'' (p. 2). \ He refers in the
preface to ``a crack in the very foundations of existing science,'' ``new
ideas and new methods that ultimately depend very little on what has gone
before,'' and ``a vast array of applications---both conceptual and
practical---that can now be developed.'' \ Comments of this character pervade
the book.

Significantly, there is no bibliography. \ Instead there are $349$ pages of
endnotes, which summarize the history, from antiquity to the present, of each
subject that Wolfram addresses. \ The notes are fascinating; in many respects
they constitute a better book than the main text. \ However, in both the main
text and in the notes, Wolfram generally brings up prior work only to dismiss
it as misguided, or at best as irrelevant to his concerns. \ For example,
after relating his `discovery' that there is apparent complexity in the
distribution of primes, Wolfram acknowledges that ``the first few hundred
primes were no doubt known even in antiquity, and it must have been evident
that there was at least some complexity in their distribution'' (p. 134).

\begin{quote}
However [he continues], without the whole intellectual structure that I have
developed in this book, the implications of this observation---and its
potential connection, for example, to phenomena in nature---were not
recognized. \ And even though there has been a vast amount of mathematical
work done on the sequence of primes over the course of many centuries, almost
without exception it has been concerned not with basic issues of complexity
but instead with trying to find specific kinds of regularities (p. 134).
\end{quote}

We believe that Wolfram is overstating his case. \ In the remainder of this
review we will explain why we believe this, by examining various specific
claims that he makes. \ In Section 2, we address some of
Wolfram's conjectures about computational complexity using standard techniques
in theoretical computer science. \ We also argue that his Principle of
Computational Equivalence does not have the relevance to $\mathsf{NP}%
$-completeness that he asserts for it.\ \ In Section 3, we review
Wolfram's ideas regarding fundamental physics, pointing out their similarity
to existing work in loop quantum gravity. \ We also examine Wolfram's proposal
for a deterministic model underlying quantum mechanics, with `long-range
threads' to connect entangled particles. \ We show that this proposal cannot
be made compatible with both special relativity and Bell inequality violations.

\vspace*{1pt}\textlineskip
\section{Computational Complexity}\label{complexity}
\vspace*{-0.5pt}
\noindent

In the opening chapter we are promised that Wolfram's new kind of science
\textquotedblleft begins to shed new light on various longstanding questions
in computational complexity theory\textquotedblright\ (p. 14). \ On pages
758--764 we learn what this means. \ Complexity theorists have struggled for
decades to prove lower bounds---for example, \textquotedblleft any Turing
machine that decides whether a Boolean formula of size $n$ is satisfiable
requires a number of steps exponential in $n$.\textquotedblright\ \ What
Wolfram proposes is restricting attention to `simple'\ Turing machines: say,
all those with $4$ states and a $2$-symbol alphabet. \ The number of such
machines is finite, so one could try to analyze all of them (with the aid of a
computer), and show that not one solves a given problem within a specified
time bound.

Supposing we did this, what would it tell us? \ Can such simple Turing
machines display nontrivial behavior? \ To demonstrate that they can, Wolfram
exhibits a machine (call it $M$) that solves a problem $L_{M}$\ in time
exponential in the length of its input. \ No machine with at most $4$ states
can solve $L_{M}$\ more efficiently, and Wolfram conjectures that $M$ is
`irreducible,'\ meaning that no machine with \textit{any} number of states can
solve $L_{M}$\ substantially more efficiently.

However, Wolfram's enumeration approach has a crucial drawback, arising from
the same phenomenon of complexity it seeks to address. \ Not only does the
number of $n$-state Turing machines grow exponentially in $n$; but analyzing
any one machine is equivalent to the `halting problem,'\ well known to be
undecidable with any amount of resources. \ More concretely, we doubt the
enumeration approach scales even to $5$-state, $2$-symbol Turing machines.
\ For as noted by Wolfram on p. 889, it is not even known what is the maximum
number of steps such a machine could make when started on a blank tape. \ (The
best known lower bound for this number, called the `$5^{th}$ Busy Beaver shift
number,' is $47,176,870$,\ due to Marxen and Buntrock \cite{mb}.)

But if extremely short programs can produce `irreducibly complex' behavior,
then aren't they already of interest? \ Indeed, in Chapter 11 Wolfram gives a
remarkable construction, due to his employee Matthew Cook, showing that even
the Rule $110$ cellular automaton is a universal computer. \ A corollary is
that there exists a $2$-state, $5$-symbol Turing machine that can simulate any
other Turing machine. \ For this construction to work, though, the Turing
machine being simulated must be encoded onto the input tape in an extremely
complicated way. \ All `tiny' universal machines known to date face the same problem.

So the real question is: given a problem of interest, such as matrix
multiplication, does an extremely short program exist to solve it using a
\textit{standard} encoding? \ We are willing to be surprised by such a
program, but Wolfram never comes close to providing an example. \ If there are
no such examples, then by restricting to extremely short programs, we are
merely trading an infeasible search among programs for an infeasible search
among input encoding schemes.

Elsewhere in the book, there are a few errors and oversights regarding
complexity. \ Wolfram says that minimizing a DNF expression (p. 1096) and
computing a permanent (p. 1146) are \textsf{NP}-complete; they are
respectively $\mathsf{\Sigma}_{2}^{P}$-complete (as shown by Umans
\cite{umans}) and $\mathsf{\#P}$-complete (as shown by Valiant \cite{valiant}%
). \ Also, in Chapter 10, pseudorandom number generators based on cellular
automata are proposed. \ Wolfram suggests that, since certain questions
involving cellular automata are \textsf{NP}-complete, these generators might
be a good basis for cryptography:

\begin{quote}
To date [no] system has been devised whose cryptanalysis is known to be
NP-complete. \ Indeed, essentially the only problem on which cryptography
systems have so far successfully been based is factoring of integers. \ [And]
while this problem has resisted a fair number of attempts at solution, it is
not known to be NP-complete (and indeed its ability to be solved in polynomial
time on a formal quantum computer may suggest that it is not) (p. 1089--1090).
\end{quote}

The most common cryptanalysis problems, such as inverting a one-way
permutation, are in $\mathsf{NP\cap coNP}$, which means that they cannot be
\textsf{NP}-complete\ unless $\mathsf{NP=coNP}$. \ On the other hand, Canetti
et al. \cite{cdno}\ have proposed `deniable encryption' schemes, in which a
single ciphertext can correspond to many plaintexts.\ \ Such a scheme could
conceivably be secure assuming only $\mathsf{P\neq NP}$, but finding a scheme
for which that implication provably holds remains a difficult open problem.

What Wolfram has proposed is simply a candidate pseudorandom generator.
\ There is no shortage of these (or equivalently, candidate one-way functions;
H\aa stad et al. \cite{hill}\ showed that either can be obtained from the
other). \ Some such generators are based on $\mathsf{NP}$-complete problems,
but that is not considered evidence for their security, since breaking the
generator might be easier than solving the $\mathsf{NP}$-complete problem.
\ Attention has focused on factoring---and on other `structured'\ problems,
involving elliptic curves, error-correcting codes, lattices, and so
on---because of the need for \textit{trapdoor} one-way functions in public-key cryptography.

\vspace*{1pt}\textlineskip
\subsection{The principle of computational equivalence}\label{principle}
\vspace*{-0.5pt}
\noindent

The final chapter proposes a `Principle of Computational Equivalence': that
almost all systems that are not `obviously simple' are in some sense
equivalent to a universal Turing machine. \ Wolfram\ emphasizes that this
principle goes beyond the Church-Turing thesis in two ways: it asserts, first,
that universality is pervasive in nature; and second, that universality arises
in any sufficiently complex system, without needing to be `engineered in.'
\ To us, the principle still seems an expression of the conventional
wisdom\ in theoretical computer science. \ However, we will not debate this
question in general terms. \ Instead we will consider a specific implication
that Wolfram offers for computational complexity:

\begin{quote}
In studying the phenomenon of NP completeness what has mostly been done in the
past is to try to construct particular instances of rather general problems
that exhibit equivalence to other problems. \ But almost always what is
actually constructed is quite complicated---and certainly not something one
would expect to occur at all often. \ Yet on the basis of intuition from the
Principle of Computational Equivalence I strongly suspect that in most cases
there are already quite simple instances of general NP-complete problems that
are just as difficult as any NP-complete problem. \ And so, for example, I
suspect that it does not take a cellular automaton nearly as complicated as
[one with 19 colors given previously] for it to be an NP-complete problem to
determine whether initial conditions exist that lead to particular behavior.
(p. 769)
\end{quote}

In computer science, the complexity of `typical' instances of $\mathsf{NP}%
$-complete problems has been investigated for decades. \ Highlights include
Levin's theory of average-case completeness \cite{levin}\ and studies of phase
transitions in randomly generated combinatorial problems \cite{cm}. \ It
remains open to show that some $\mathsf{NP}$-complete problem is hard on
average, under a simple distribution, so long as $\mathsf{P}\neq\mathsf{NP}$.
\ However, `worst-case/average-case equivalence' has been shown for several
cryptographic problems,\ including one studied by Ajtai and Dwork \cite{ad}.

As for the cellular automaton conjecture, its validity depends on how it is
formulated. \ Suppose we are given a one-dimensional, two-color cellular
automaton on a lattice of bounded size $n$, and an\ ending condition
$E\in\left\{  0,1\right\}  ^{n}$.\ \ Then we can decide in polynomial time
whether there exists an initial condition that evolves to $E$ in one or more
steps, by using dynamic programming. \ Extending this technique, we can decide
whether there exists an initial condition that evolves to $E$ in exactly $t$
steps, where $t=O\left(  \log n\right)  $, by computing a list of all possible
initial configurations for each contiguous block of $t$\ cells.

Indeed, for any fixed polynomial-time predicate $\Phi$, let
$Init_{110}^{\Phi}$\ be the following problem. \ We are given
an ending condition $E$ with $n$ cells, and asked to decide whether there
exists an initial condition $I$ such that (i) $\Phi\left(  I\right)  $\ holds,
and (ii) the Rule $110$ cellular automaton evolves $I$ to $E$ in exactly $t$
steps.\ \ Here $t$ is a fixed polynomial in $n$. \ Then:

\begin{proposition}
For all $\Phi$\ and polynomials $p$, there is a polynomial-time algorithm that
solves a $1-1/p\left(  n\right)  $\ fraction of $Init%
_{110}^{\Phi}$\ instances.
\end{proposition}

\begin{proof}
Consider a directed graph with $2^{n}$ vertices, one for each configuration,
and edges corresponding to the action of Rule $110$. \ Each vertex has
outdegree $1$, so the number of paths of length $t$ is $2^{n}$. \ Thus, if $E
$ is chosen uniformly at random, then the expected number of length-$t$ paths
ending at $E$ is $1$, so this number is at most $p\left(  n\right)  $\ with
probability at least $1-1/p\left(  n\right)  $. \ In this case we can trace
$t$ steps backward from $E$ in time $O\left(  ntp\left(  n\right)  \right)  $,
maintaining a list of all possible predecessors, and then evaluate $\Phi$\ on each.
\end{proof}

Nevertheless, since Rule $110$ is universal, the following intuition suggests
that $Init_{110}^{\Phi}$\ should be $\mathsf{NP}%
$-complete\ for\ some $\Phi$. \ Given a Boolean formula $\Psi$\ and proposed
solution $X$, we could create an initial condition $I\left(  \Psi,X\right)  $
corresponding to a Turing machine that checks whether $X$ satisfies $\Psi$;
and if it does, erases $X$, preserves\ $\Psi$, and goes into an
`accept'\ state $A\left(  \Psi\right)  $. \ Then by having $\Phi$\ verify that
the initial condition is of legal form, we could reduce the problem of whether
$\Psi$\ is satisfiable to that of whether there exists a legal initial
condition that evolves to\ $E=A\left(  \Psi\right)  $.

This intuition fails for an interesting reason. \ Cook's proof that Rule $110$
is universal relies on simulating `cyclic tag systems,' a variant of the tag
systems studied in the 1960's by Cocke and Minsky \cite{cm0}\ among others
(see also p. 670 of Wolfram). \ However, though Wolfram does not discuss this
explicitly in the book, the known simulations of Turing machines by tag
systems require exponential slowdown. \ To prove that $Init%
_{110}^{\Phi}$\ is $\mathsf{NP}$-complete, what is needed is to show that Rule
$110$ allows \textit{efficient} simulation of Turing machines.

In summary, there are many fascinating complexity questions about
one-dimensional cellular automata, as well as about typical instances of
$\mathsf{NP}$-complete problems. \ But it is unclear why the Principle of
Computational Equivalence should yield more insight into these questions than
the standard techniques of computational complexity.

\vspace*{1pt}\textlineskip
\section{Fundamental Physics}\label{physics}
\vspace*{-0.5pt}
\noindent

The most interesting chapter of \textit{A New Kind of Science} is the ninth,
on `Fundamental Physics.' \ Here Wolfram confronts general relativity and
quantum mechanics, arguably the two most serious challenges to a view of
nature based on deterministic cellular automata. \ He conjectures that
spacetime is discrete at the Planck scale, of about $10^{-33}$\ centimeters or
$10^{-43}$\ seconds. \ This conjecture is not new; it has long been considered
in the context of loop quantum gravity \cite{ms,rs}, and has also received
attention in connection with the holographic principle \cite{bousso}\ from
black hole thermodynamics. \ But are new ideas offered to substantiate the conjecture?

For Wolfram, spacetime is a causal network, in which events are vertices and
edges specify the dependence relations between events. \ Pages 486--496 and
508--515 discuss in detail how to generate such a network from a simple set of
rules. \ In particular, we could start with a finite undirected `space graph'
$G$, assumed to be regular with degree $3$ (since higher-degree vertices can
be replaced by cycles of degree-$3$ vertices). \ We then posit a set of update
rules, each of which replaces a subgraph by another subgraph with the same
number of outgoing edges. \ The new subgraph must preserve any symmetries of
the old one. \ Then each event in the causal network corresponds to an
application of an update rule. \ If updating event $B$ becomes possible as a
result of event $A$, then we draw an edge from $A$ to $B$.

Properties of space are defined in terms of $G$. \ For example, if the number
of vertices in $G$ at distance at most $n$ from any given vertex grows as
$n^{D}$, then space can be said to have dimension $D$. \ (As for formalizing
this definition, Wolfram says only that there are \textquotedblleft some
subtleties. \ For example, to find a definite volume growth rate one does
still need to take some kind of limit---and one needs to avoid sampling too
many or too few\textquotedblright\ vertices \ (p. 1030).) \ Similarly, Wolfram
asserts that the curvature information needed for general relativity, in
particular the Ricci tensor, can be read from the connectivity pattern of $G$.
\ To make the model as simple as possible, Wolfram does not associate a bit to
each vertex of $G$, representing (say) the presence or absence of a particle.
\ Instead particles are localized structures, or `tangles,' in $G$.

The above ideas have all been discussed previously by researchers in quantum
gravity: in particular, that spacetime is a causal network arising from graph
updating rules \cite{ms}; that particles could arise as `topological defects'
in such a network \cite{crane}; and that dimension and other geometric
properties can be defined solely in terms of the network's connectivity
pattern \cite{nr}. \ The main difference we can discern between Wolfram's
model and earlier ones is that Wolfram's is explicitly \textit{classical}.
\ Indeed, Wolfram requires the network evolution to be deterministic, by
disallowing `multiway systems': that is, sets of update rules that can yield
nonequivalent causal networks, depending on the order in which rules are
applied. \ He opts instead for rule sets that are `causal invariant,' i.e.
that yield the same network regardless of rule application order. \ As noted
by Wolfram, a sufficient (though not necessary) condition for causal
invariance is that no `replaceable' subgraph overlaps itself or any other
replaceable subgraph.

Wolfram points out an immediate analogy to special relativity, wherein
observers do not in general agree on the order in which spacelike separated
events occur, yet agree on any final outcome of the events. \ He is vague,
though, about how (say) the Lorentz transformations might be derived:

\begin{quote}
There are many subtleties here, and indeed to explain the details of what is
going on will no doubt require quite a few new and rather abstract concepts.
But the general picture that I believe will emerge is that when particles move
faster they will appear to have more nodes associated with them (p. 529).
\end{quote}

Wolfram is \textquotedblleft certainly aware that many physicists will want to
know more details,\textquotedblright\ he writes in the endnotes, about how his
model can reproduce known features of physics. \ But, although he chose to
omit technical formalism from the presentation, \textquotedblleft\lbrack
g]iven my own personal background in theoretical physics it will come as no
surprise that I have often used such formalism in the process of working out
what I describe in these sections\textquotedblright\ (p. 1043). \ The paradox
is obvious: if technical formalism would clarify his ideas, then what could
Wolfram lose by including it in the endnotes? \ If, on the other hand, such
formalism is irrelevant, then why does Wolfram even mention having used it?

\vspace*{1pt}\textlineskip
\subsection{Quantum mechanics}\label{quantum}
\vspace*{-0.5pt}
\noindent

Physicists' hunger for details will likely grow further when they read the
section on `Quantum Phenomena' (p. 537--545). \ Here Wolfram maintains that
quantum mechanics is only an approximation to an underlying classical (and
most likely deterministic) theory. \ Many physicists have sought such a
theory, from Einstein to (in modern times) 't Hooft \cite{thooft}. \ But a
series of results, beginning in the 1960's, has made it clear that such a
theory comes at a price. \ Although Wolfram discusses these results, in our
view he has not understood what they entail. \ Because this point is an
important one, we will devote this section and the next to it.

To begin, Wolfram is \textit{not} advocating a hidden-variable approach such
as Bohmian mechanics, in which the state vector is supplemented by an
`actual'\ eigenstate of a particular observable. \ Instead he thinks that, at
the lowest level, the state vector is not needed at all; it is merely a useful
construct for describing some (though presumably not all) higher-level
phenomena. \ Indeterminacy arises because of one's inability to know the exact
state of a system:

\begin{quote}
[I]f one knew all of the underlying details of the network that makes up our
universe, it should always be possible to work out the result of any
measurement. \ I strongly believe that the initial conditions for the universe
were quite simple. \ But like many of the processes we have seen in this book,
the evolution of the universe no doubt intrinsically generates apparent
randomness. \ And the result is that most aspects of the network that
represents the current state of our universe will seem essentially random (p. 543).
\end{quote}

Similarly, Wolfram explains as follows why an electron has wave properties:
``\ldots a network which represents our whole universe must also include us as
observers. \ And this means that there is no way that we can look at the
network from the outside and see the electron as a definite object'' (p. 538).
\ An obvious question then is how Wolfram accounts for the possibility of
quantum computing, assuming $\mathsf{BPP}\neq\mathsf{BQP}$. \ He gives an
answer in the final chapter:

\begin{quote}
Indeed within the usual formalism [of quantum mechanics] one can construct
quantum computers that may be able to solve at least a few specific problems
exponentially faster than ordinary Turing machines. \ But particularly after
my discoveries in Chapter 9 [`Fundamental Physics'], I strongly suspect that
even if this is formally the case, it will still not turn out to be a true
representation of ultimate physical reality, but will instead just be found to
reflect various idealizations made in the models used so far (p. 771).
\end{quote}

In the endnotes, though, where he explains quantum computing in more detail,
Wolfram seems to hedge about which idealizations he has in mind:

\begin{quote}
It does appear that only modest precision is needed for the initial
amplitudes. \ And it seems that perturbations from the environment can be
overcome using versions of error-correcting codes. \ But it remains unclear
just what might be needed actually to perform for example the final
measurements required (p. 1148).
\end{quote}

One might respond that, with or without quantum computing, Wolfram's proposals
can be ruled out on the simpler ground that they disallow Bell inequality
violations. \ However, Wolfram puts forward an imaginative hypothesis to
account for bipartite entanglement. \ When two particles (or `tangles' in the
graph $G$) collide, long-range `threads' may form between them, which remain
in place even if the particles are later separated:

\begin{quote}
The picture that emerges is then of a background containing a very large
number of connections that maintain an approximation to three-dimensional
space, together with a few threads that in effect go outside of that space to
make direct connections between particles (p. 544).
\end{quote}

The threads can produce Bell correlations, but are somehow too small (i.e.
contain too few edges) to transmit information in a way that violates
causality. \ This is reminiscent of, for example, the
`multisimultaneity'\ model studied experimentally by Stefanov et al.
\cite{szgs}.

There are several objections one could raise against this thread hypothesis.
\ What we will show in Section 3.2 is that, \textit{if} one accepts
three of Wolfram's own desiderata---determinism, relativity of inertia, and
causal invariance---then the hypothesis fails. \ For now, though, we remark
that Wolfram says little about what, to us, is a more natural possibility than
the thread hypothesis. \ This is an explicitly \textit{quantum} cellular
automaton or causal network, with a unitary transition rule. \ (For
discussions of how to construct such automata, see van Dam \cite{vandam}\ and
Watrous \cite{watrous}.) \ The reason seems to be that he does not want
continuity anywhere in a model, not even in probabilities or amplitudes. \ In
the notes, he describes an experiment with a quantum cellular automaton as follows:

\begin{quote}
One might hope to be able to get an ordinary cellular automaton with a limited
set of possible values by choosing a suitable [phase rotation] $\theta$
[$\theta=\pi/4$ and $\theta=\pi/3$\ are given as examples in an illustration].
\ But in fact in non-trivial cases most of the cells generated at each step
end up having distinct values (p. 1060).
\end{quote}

This observation is unsurprising, given results in quantum computing to the
effect that almost any nontrivial gate set is universal (that is, can
approximate any unitary matrix to any desired precision, or any orthogonal
matrix in case one is limited to reals). \ Indeed, Shi \cite{shi}\ has shown
that a Toffoli gate\footnote{Indeed one can use almost any classical
reversible $3$-bit gate in place of a Toffoli gate.\ \ On p. 1098, Wolfram
reports that, out of $40,320$ such classical reversible gates, $38,976$ are
universal.} plus any gate that does not preserve the computational basis, or a
controlled-NOT gate plus any gate whose \textit{square} does
not preserve the computational basis, are both universal gate sets. \ In any
case, Wolfram does not address the fact that continuity in amplitudes seems
more `benign' than continuity in measurable quantities: the former, unlike the
latter, does not enable an infinite amount of computation to be performed in a
finite time. \ Also, as observed by Bernstein and Vazirani \cite{bv}, the
linearity of quantum mechanics implies that small errors in amplitudes cannot
be magnified during a quantum computation.

\vspace*{1pt}\textlineskip
\subsection{Bell's theorem and causal invariance}\label{bell}
\vspace*{-0.5pt}
\noindent

Our goal in this section is to show that Wolfram's\ `long-range thread'\ model
cannot be made compatible with both special relativity and Bell inequality
violations. \ Moreover, this is not a fixable oversight, but a basic
shortcoming of any such model. \ One might think this conclusion immediate.
\ Wolfram, however, allows two key ingredients: first, nonlocality, in the
form of the long-range threads; and second, `randomness,' in the form of
observers' inability to know the complete state of the causal network (the
latter is discussed on pages 299--326). \ We will argue that these ingredients
do not suffice. \ In any model of the sort Wolfram considers, randomness must
play a more fundamental role than he allows\footnote{A similar argument could
be made on the basis of the Kochen-Specker theorem \cite{held}. \ The reason
we have not done so is that Wolfram never explicitly requires his model to be
noncontextual.}.

We now make the argument more formal. \ Let $\mathcal{R}$ be a set of graph
updating rules, which might be probabilistic. \ Then\ consider the following
four assertions (which, though not mathematically precise, will be clarified
by subsequent discussion).

\begin{enumerate}
\item[(1)] $\mathcal{R}$ satisfies causal invariance. \ That is, given any
initial graph (and choice of randomness if $\mathcal{R}$\ is probabilistic),
$\mathcal{R}$\ yields a unique causal network.

\item[(2)] $\mathcal{R}$ satisfies the relativity postulate. \ That is,
assuming the causal network approximates a flat Minkowski spacetime at a large
enough scale, there are no preferred inertial frames.

\item[(3)] $\mathcal{R}$ permits Bell inequality violations.

\item[(4)] Any updating rule in $\mathcal{R}$ is always considered to act on a
fixed graph, not on a distribution or superposition over graphs. \ This is
true even if parts of the initial graph are chosen at random, and even if
$\mathcal{R}$ is probabilistic.
\end{enumerate}

Our goal is to show that, for any $\mathcal{R}$, at least one of these
assertions is false. \ Current physical theory would suggest that (1)-(3) are
true and that (4) is false. \ Wolfram, if we understand him correctly, starts
with (4) as a premise, and then introduces causal invariance to satisfy (1)
and (2), and long-range threads to satisfy (3).

In a standard Bell experiment, Alice and Bob are given input bits $x_{A}$\ and
$x_{B}$\ respectively, chosen uniformly and independently at random. \ Their
goal is, without communicating, to output bits $y_{A}$\ and $y_{B}$
respectively such that $y_{A}\oplus y_{B}=x_{A}\wedge x_{B}$. \ Under any
`local hidden variable' theory, Alice and Bob can succeed with probability at
most $3/4$; the optimal strategy is for them to ignore their inputs and output
(say) $y_{A}=0$\ and $y_{B}=0$. \ However, suppose Alice has a qubit $\rho
_{A}$\ and Bob a $\rho_{B}$, that are jointly in the Bell state $\left(
\left\vert 00\right\rangle +\left\vert 11\right\rangle \right)  /\sqrt{2}$.
\ Then there is a protocol\footnote{If $x_{A}=1$\ then Alice applies a $\pi
/8$\ phase rotation to $\rho_{A}$, and if $x_{B}=1$\ then Bob applies a
$-\pi/8$\ rotation to $\rho_{B}$. \ Both parties then measure in the standard
basis and output whatever they observe.} by which they can succeed with
probability $\left(  5+\sqrt{2}\right)  /8\approx0.802$.

We model this situation by letting $A$ and $B$, corresponding to Alice and
Bob, be disjoint subgraphs of a graph $G$. \ We suppose that, at a large
scale, $G$ approximates a Euclidean space of some dimension; and that any
causal network obtained by applying updates to $G$\ approximates a Minkowski
spacetime. \ We can think of $G$ as containing long-range threads from $A$ to
$B$, though the nature of the threads will not affect our conclusions. \ We
encode Alice's input $x_{A}$\ by (say) placing an edge between two specific
vertices in $A$ if and only if $x_{A}=1$. \ We encode $x_{B}$ similarly, and
also supply Alice and Bob with arbitrarily many correlated random bits.
\ Finally we stipulate that, at the end of the protocol, there is an edge
between two specific vertices in $A$ if and only if $y_{A}=1$, and similarly
for $y_{B}$. \ A technicality is that we need to be able to identify which
vertices correspond to $x_{A}$, $y_{A}$, and so on, even as $G$ evolves over
time. \ We could do this by stipulating that (say) \textquotedblleft the
$x_{A}$\ vertices are the ones that are roots of complete binary trees of
depth $3$,\textquotedblright\ and\ then choosing the rule set to guarantee
that, throughout the protocol, exactly two vertices have this property.

Call a variable `touched' after an update has been applied to a subgraph
containing any of the variable's vertices.\ \ Also, let $Z$ be an assignment
to all random variables: that is, $x_{A}$, $x_{B}$, the correlated random
bits, and the choice of randomness if $\mathcal{R}$\ is probabilistic. \ Then
for all $Z$ we require the following, based on what observers in different
inertial frames could perceive:

\begin{enumerate}
\item[(i)] There exists a sequence of updates under which $y_{A}$\ is output
before any of Bob's variables are touched.

\item[(ii)] There exists another sequence under which $y_{B}$\ is output
before any of Alice's variables are touched.
\end{enumerate}

Then it is easy to see that, if a Bell inequality violation occurs, then
causal invariance must be violated. \ Given $Z$, let $y_{A}^{\left(  1\right)
}\left(  Z\right)  $, $y_{B}^{\left(  1\right)  }\left(  Z\right)  $ be the
values of $y_{A},y_{B}$\ that are output under rule sequence (i), and let
$y_{A}^{\left(  2\right)  }\left(  Z\right)  $, $y_{B}^{\left(  2\right)
}\left(  Z\right)  $\ be the values output under sequence (ii). \ Then there
must exist some $Z$ for which either $y_{A}^{\left(  1\right)  }\left(
Z\right)  \neq y_{A}^{\left(  2\right)  }\left(  Z\right)  $ or $y_{B}%
^{\left(  1\right)  }\left(  Z\right)  \neq y_{B}^{\left(  2\right)  }\left(
Z\right)  $---for if not, then the entire protocol could be simulated under a
local hidden variable model. \ It follows that the outcome of the protocol can
depend on the order in which updates are applied.

To obtain a Bell inequality violation, something like the following seems to
be needed. \ We can encode `hidden variables'\ into $G$, representing the
outcomes of the possible measurements Bob could make on $\rho_{B}$. \ (We can
imagine, if we like, that the update rules are such that observing any one of
these variables destroys all the others. \ Also, we make no assumption of
contextuality.) \ Then, after Alice measures $\rho_{A}$, using the long-range
threads she updates Bob's hidden variables conditioned on her measurement
outcome. \ Similarly, Bob updates Alice's hidden variables conditioned on his
outcome. \ Since at least one party must access its hidden variables for there
to be a Bell inequality violation, causal invariance is still violated. \ But
a sort of probabilistic causal invariance holds, in the sense that\ if we
marginalize out $A$ (the `Alice' part of $G$), then the \textit{distribution}
of values for each of Bob's hidden variables is the same before and after
Alice's update.\ \ The lesson is that, if we want both causal invariance and
Bell inequality violations, then we need to introduce probabilities at a
fundamental level---not merely to represent Alice and Bob's subjective
uncertainty about the state of $G$, but even to define whether a set of rules
is or is not causal invariant.

Note that we made no assumption about how the random bits were
generated---i.e. whether they were `truly random' or were the pseudorandom
output of some updating rule.\ \ Our conclusion is also unaffected if we
consider a `deterministic' variant of Bell's theorem due to Greenberger,
Horne, and Zeilinger \cite{ghz}, discussed by Wolfram on p. 1065. \ There
three parties, Alice, Bob, and Charlie, are given input bits $x_{A}$, $x_{B}$,
and $x_{C}$ respectively, satisfying the promise that $x_{A}\oplus x_{B}\oplus
x_{C}=0$. \ The goal is to output bits $y_{A}$, $y_{B}$, and $y_{C}$\ such
that $y_{A}\oplus y_{B}\oplus y_{C}=x_{A}\vee x_{B}\vee x_{C}$. \ Under a
local hidden variable model, there is no protocol that succeeds on all four
possible inputs; but if the parties share the GHZ state $\left(  \left\vert
011\right\rangle +\left\vert 101\right\rangle +\left\vert 110\right\rangle
-\left\vert 000\right\rangle \right)  /2$, then such a protocol exists.
\ However, although the \textit{output} is correct with certainty, assuming
causal invariance one cannot \textit{implement} the protocol in the long-range
thread model, for precisely the same reason as for the two-party Bell
inequality. \ Again, one needs to be able to talk about the state of $G$ as a
distribution or superposition over classical states.

After Wolfram was sent a version of this review, Rowland \cite{rowland}, an
employee of Wolfram,\ wrote to us that the above argument fails for the
following reason. \ We assumed that there exist two sequences of updating
events, one in which Alice's measurement precedes Bob's and one in which Bob's
precedes Alice's. \ But we neglected the possibility that a \textit{single}
update, call it $E$, is applied to a subgraph that straddles the long-range
threads. \ The event $E$ would encompass both Alice and Bob's measurements, so
that neither would precede the other in any sequence of updates. \ We could
thereby obtain a rule set $\mathcal{R}$\ satisfying assertions (1), (3), and (4).

We argue that such an $\mathcal{R}$\ would nevertheless fail to satisfy (2).
\ For in effect we start with a flat Minkowski spacetime, and then take two
distinct events that are simultaneous in a particular inertial frame, and
identify them as being the \textit{same} event $E$. \ This can be visualized
as `pinching together' two horizontally separated points on a spacetime
diagram. \ (Actually a whole `V' of points must be pinched together, since
otherwise entanglement could not have been created.) \ However, what happens
in a different inertial frame? \ It would seem that $E$, a single event, is
perceived to occur at two separate times. \ That by itself might be thought
acceptable, but it implies that there exists a class of preferred inertial
frames: those in which $E$ is perceived to occur only once. \ Of course, even
in a flat spacetime, one could designate as `preferred'\ those frames in which
Alice and Bob's measurements are perceived to be simultaneous. \ A crucial
distinction, though, is that there one only obtains a class of preferred
frames after deciding which event at Alice's location, \textit{and} which
event at Bob's location, should count as `measurements.' \ Under Rowland's
hypothesis, by contrast, once one decides what counts as the measurement at
Alice's location, the decision at Bob's location is made automatically. \ This
is because the measurement involves $E$, an event which straddles the two
locations (which would otherwise be spacelike separated).

\vspace*{1pt}\textlineskip
\section{Conclusion}
\vspace*{-0.5pt}
\noindent

Steven Levy \cite{levy}\ opines in \textit{Wired}\ that ``probably the
toughest criticism [of \textit{A New Kind of Science}] will come from those
who reject Wolfram's ideas because the evidence for his contentions is based
on observing systems contained inside computers.'' \ In our opinion, however,
it is preferable to judge the book on its own terms---to grant, that is, that
many complex systems might indeed be fruitfully understood in terms of simple
computations. \ The question is, what does the book tell us about such
systems, beyond what was known from `traditional science'?

This review focused on two fields, computational complexity and fundamental
physics, which Wolfram claims are transformed by the discoveries in his book.
\ We made no attempt to assess the book's relevance to other fields such as
evolutionary biology and statistical physics.

In computational complexity, we argued that Wolfram tends to recapitulate
existing ideas (such as pseudorandomness and the intractability of simple
instances of \textsf{NP}-complete problems), albeit without precise
definitions or proofs, and with greater claims of significance. \ For
theoretical computer scientists, the most interesting content in the book will
possibly be the explicit constructions of Turing machines and cellular automata.

In physics, the book proposes that spacetime be viewed in terms of causal
networks arising from graph rewriting systems. \ We argued that this proposal,
as well as Wolfram's elaborations on it, have been previously considered in
the loop quantum gravity literature. \ Wolfram claims to have further details
to validate the proposal, but has declined to supply them. \ As for the idea
that a deterministic, relativistically invariant, causal invariant model
underlies quantum mechanics, we argued that it fails---even if quantum
mechanics breaks down for more than two particles, and even if, as Wolfram
suggests, one allows long-range threads to connect entangled particles.
\ Exactly what kinds of classical models could underlie quantum mechanics is a
question of great importance, but Wolfram makes no serious effort to address
the question.

\textit{A New Kind of Science} was published by Wolfram's own company, and was
not subject to outside editing or peer review. \ If it were (say) a
creationist tract, then this unusual route to publication would be of little
consequence: for then no amount of editing would have improved it, and few
scientifically literate readers would be misled by it. \ What is unfortunate
in this case is that outside editing would probably have made a substantial
difference. \ In an endnote, Wolfram explains that ``[p]erhaps I might avoid
some criticism by a greater display of modesty [in the text], but the cost
would be a drastic reduction in clarity'' (p. 849). \ However, were the book
more cautious in its claims and more willing to acknowledge previous work, it
would likely be easier for readers to assess what it does offer: a
cellular-automaton-based perspective on existing ideas in science.\ \ Thus, we
believe the book would be not only less susceptible to criticism, but also clearer.

\nonumsection{Acknowledgements}
\noindent
I thank Stephen Wolfram for an enjoyable conversation about this review; David
Reiss for arranging the conversation; Todd Rowland for correspondence about
Section 3.2; and Barbara Terhal, Lee Smolin, Robert Solovay, Moni Naor,
and Dave Bacon for helpful comments. \ Supported by an NSF Graduate Fellowship
and by DARPA grant F30602-01-2-0524.

\vspace{0.13in}
\noindent {\bf Scott Aaronson} (aaronson@cs.berkeley.edu) \\
Computer Science Department \\
University of California, Berkeley, CA 94720, USA

\nonumsection{References}


\noindent

\begin{thebibliography}{000}
\bibitem {feynman}R. P. Feynman (1998),\ \textit{The Character of Physical
Law}, MIT Press (originally published 1965).

\bibitem {wolfram}S. Wolfram (2002),\ \textit{A New Kind of Science}, Wolfram Media.

\bibitem {farmela}G. Farmela (13 January 1999),\ The universe in black and
white, \textit{The Daily Telegraph}.

\bibitem {mb}H. Marxen and J. Buntrock (1990),\ Attacking the Busy Beaver $5$,
\textit{Bulletin of the EATCS} 40, pp. 247--251. \ Also
www.drb.insel.de/\symbol{126}heiner/BB.

\bibitem {umans}C. Umans (1998),\ The minimum equivalent DNF problem and
shortest implicants, \textit{Proceedings of the 39th IEEE\ Symposium on
Foundations of Computer Science (FOCS)}, pp. 556--563.

\bibitem {valiant}L. G. Valiant (1979),\ The complexity of computing the
permanent, \textit{Theoretical Computer Science} 8(2), pp. 189--201.

\bibitem {cdno}R. Canetti, C. Dwork, M. Naor, and R. Ostrovsky
(1997),\ Deniable encryption, \textit{Proceedings of CRYPTO'97}.

\bibitem {hill}J. H\aa stad, R. Impagliazzo, L. A. Levin, and M. Luby
(1999),\ A pseudorandom generator from any one-way function, \textit{SIAM
Journal on Computing} 28(4), pp. 1364--1396.

\bibitem {levin}L. A. Levin (1986),\ Average case complete problems,
\textit{SIAM Journal on Computing} 15(1), pp. 285--286.

\bibitem {cm}S. A. Cook and D. Mitchell (1997),\ Finding hard instances of the
satisfiability problem: a survey, \textit{DIMACS Series in Discrete Math and
Theoretical Computer Science} 35, pp. 1--17.

\bibitem {ad}M. Ajtai and C. Dwork (1996),\ A public-key cryptosystem with
worst-case/average-case equivalence, \textit{Electronic Colloquium on
Computational Complexity (ECCC)} 3(65).

\bibitem {cm0}J. Cocke and M. Minsky (1964),\ Universality of tag systems with
$P=2$, \textit{Journal of the ACM} 11(1), pp. 15--20.

\bibitem {ms}F. Markopoulou and L. Smolin (1997),\ Causal evolution of spin
networks, \textit{Nuclear Physics} B508:409. \ gr-qc/9702025.

\bibitem {rs}D. P. Rideout and R. D. Sorkin (2000),\ A classical sequential
growth dynamics for causal sets, \textit{Physical Review} D61. \ gr-qc/9904062.

\bibitem {bousso}R. Bousso,\ The holographic principle (2002), \textit{Reviews
of Modern Physics} 74(3). \ hep-th/0203101.

\bibitem {crane}L. Crane (2001),\ A new approach to the geometrization of
matter, gr-qc/0110060.

\bibitem {nr}T. Nowotny and M. Requardt (1998),\ Dimension theory of graphs
and networks, \textit{Journal of Physics }A31, pp. 2447--2463. \ hep-th/9707082.

\bibitem {thooft}G. 't Hooft (1999),\ Quantum gravity as a dissipative
deterministic system, \textit{Classical and Quantum Gravity} 16, 3263--3279.

\bibitem {szgs}A. Stefanov, H. Zbinden, N. Gisin, and A. Suarez
(2002),\ Quantum correlations with spacelike separated beam splitters in
motion: experimental test of multisimultaneity, \textit{Physical Review
Letters} 88(12).

\bibitem {vandam}W. van Dam (1996),\ Quantum cellular automata, Master's
Thesis, University of Nijmegen, The Netherlands.
\ http://www.cs.berkeley.edu/\symbol{126}vandam/qca.ps.gz.

\bibitem {watrous}J. Watrous (1995),\ On one-dimensional quantum cellular
automata, \textit{Proceedings of the 36th IEEE Symposium on Foundations of
Computer Science (FOCS)}.

\bibitem {shi}Y. Shi (2002),\ Both Toffoli and controlled-NOT need little help
to do universal quantum computation, quant-ph/0205115.

\bibitem {bv}E. Bernstein and U. Vazirani (1997),\ Quantum complexity
theory,\ \textit{SIAM\ Journal on Computing} 26(5), pp.1411--1473.

\bibitem {held}C. Held (2000),\ The Kochen-Specker theorem, in
\textit{Stanford Encyclopedia of Philosophy}. \ http://plato.stanford.edu/entries/kochen-specker.

\bibitem {ghz}D. M. Greenberger, M. A. Horne, and A. Zeilinger (1990),\ Bell's
theorem without inequalities, in \textit{Sixty-Two Years of Uncertainty:
Historical, Philosophical, and Physical Inquiries into the Foundations of
Quantum Mechanics}, A. Miller (ed.), Plenum.

\bibitem {rowland}T. Rowland (June 2002),\ email correspondence.

\bibitem {levy}S. Levy (June 2002),\ The man who cracked the code to
everything, \textit{Wired}.
\end{thebibliography}
\end{document}